# The Future of Feedback:
# How Can AI Help Transform Feedback to Be More Engaging, Effective, and Scalable?

**Authors and Affiliations:**
Jennifer Meyer*, University of Vienna, Vienna, Austria
Olaf Köller*, Leibniz Institute for Science and Mathematics Education, Kiel, Germany
Thorben Jansen*, Leibniz Institute for Science and Mathematics Education, Kiel, Germany
Johanna Fleckenstein*, University of Hildesheim, Germany
Michael W. Asher, Carnegie Mellon University, USA
Sarah Bichler, Universität Passau, Passau, Germany
Laura Brandl, LMU Munich, Germany
Jasmin Breitwieser, DIPF | Leibniz Institute for Research and Information in Education, Frankfurt, Germany
Kai S. Cortina, University of Michigan, USA
Mutlu Cukurova, University College London, United Kingdom
Martin Daumiller, University of Freiburg, Germany
Hannah Deininger, University of Tübingen, Germany
Frank Fischer, LMU, Germany
Dragan Gašević, Monash University, Clayton, VIC, Australia
Jeanine Grütter, LMU Munich, Germany
Anna Hilz, Leibniz Institute for Science and Mathematics Education, Kiel, Germany
Ioana Jivet, CATALPA, FernUniversität in Hagen, Germany
Jelena Jovanović, University of Belgrade, Serbia
Rene F. Kizilcec, Cornell University, USA
Livia Kuklick, Humboldt Universität zu Berlin, Berlin, Germany
Marlit Annalena Lindner, Leibniz Institute for Science and Mathematics Education & Europa-Universität Flensburg, Germany
Anastasiya Lipnevich, NBME & the Graduate Center, City University of New York, USA
Ute Mertens, Leibniz Institute for Science and Mathematics Education, Kiel, Germany
Detmar Meurers, Leibniz Institut für Wissensmedien, Tübingen, Germany
Kou Murayama, University of Tübingen, Germany
Tanya Nazaretsky, EPFL, Switzerland
Knut Neumann, Leibniz Institute for Science and Mathematics Education, Kiel, Germany
Ernesto Panadero, Dublin City University & Deusto University
Maciej Pankiewicz, University of Pennsylvania, USA
Zachary A. Pardos, UC Berkeley, USA
Chris Piech, Stanford University, USA
Hannah Pünjer, Leibniz Institute for Science and Mathematics Education, Kiel, Germany
Nikol Rummel, Ruhr-University-Bochum and CAIS, Bochum, Germany
Marlene Steinbach, Leibniz Institute for Science and Mathematics Education, Kiel, Germany
Olga Viberg, KTH Royal Institute of Technology, Sweden
Naomi Winstone, University of Surrey, UK

 * Equal contribution

**Corresponding author:**
Jennifer Meyer, jennifer.meyer@univie.ac.at

# The Future of Feedback:
# How Can AI Help Transform Feedback to Be More Engaging, Effective, and Scalable?

**Abstract**
With digital learning environments becoming more prevalent, the ease with which generative AI enables the scalable production of real-time, automated feedback holds the potential to reshape learning and teaching experiences. This meeting report synthesizes the interdisciplinary perspectives of 50 scholars from educational psychology, computer science, science education, and the learning sciences on the use of generative AI for feedback and its promises and risks in educational practice. We highlight points of convergence in the scholarship, identify areas of debate and unresolved challenges, and outline open questions and future directions for research and educational practice that emerged from structured small-group activities designed to bridge disciplinary barriers.

**Introduction**

The purpose of educational feedback is to support and improve learning by providing learners with information and opportunities for dialogue about their current performance in relation to an academic goal, and guidance on how to close the gap between where they are and where they need to be to adapt their thinking or behavior[1-4]. Advances in generative artificial intelligence (e.g., GenAI) offer a potential solution for enabling more frequent, task-specific, personalized, and even dialogic feedback that can foster learning and motivation, thereby reshaping educational practice and research[5-9]. Although a large body of literature has examined how to provide effective feedback, results are heterogeneous, with no clear consensus about the determinants of effective feedback across educational contexts[10]. Open questions concern how to design feedback with AI to support dialogic and agentic student engagement, how to integrate it into classrooms, and how to evaluate it to ensure its effectiveness for all students at scale.

To discuss these questions, 50 established and early-career researchers from around the world, representing multiple disciplines, including computer science, learning analytics, educational psychology, the learning sciences, computational linguistics, and discipline-specific education, gathered at the "Future of Feedback Symposium", a two-day workshop in Marbach Castle, Germany, in June 2025. Through interactive small-group activities, interdisciplinary teams engaged with findings from the feedback literature, identifying remaining research gaps, with a particular focus on the role of AI, not only in solving current challenges but also in creating new ones, and how these could be mitigated.

In the following, we present three key topics (see Figure 1 for an overview) that emerged from these activities, highlighting insights and discussions on (1) determinants of feedback effectiveness, drawing on the broader feedback literature; (2) the role of feedback engagement, perception, and use; and (3) the challenges of implementing feedback at scale. For each of these topics, we discussed the role of AI in shaping the future of feedback. We conclude with a call for future research.

(1) **Determinants of Feedback Effectiveness**

The theoretical foundation of feedback in educational contexts was discussed, pointing to seminal works in the field[1,11-14]. With some frameworks using different terms but targeting similar processes, some participants called for more theoretical and methodological integration across feedback frameworks, highlighting the diversity of frameworks in the field[12] and the corresponding fragmented evidence. Such an interdisciplinary framework would need to integrate cognitive, social-emotional, educational, ethical, subject-specific, and computational insights with practical design principles, also considering the specificities of AI-enhanced feedback. However, the value of integration of frameworks depends on the degree of overlap and specificity among theories. Other participants therefore highlighted the need for diverse frameworks in the field, as well as the need for additional theory building that would allow making testable assumptions[15].

Workshop participants identified seminal work on feedback that established principles of effective feedback. Meta-analyses show that elaborated feedback (e.g., explanations, hints, or worked examples) outperforms simpler feedback variants (i.e., informing learners whether their answer was correct or what the correct answer was) across different learning outcomes[1-2, 10, 16-22]. Yet simpler feedback, such as score-based feedback, can also be effective, for example, in the context of automated essay evaluation[19]. Despite the availability of many meta-analyses on feedback, some participants cautioned against overstating the conclusiveness of their results, suggesting that the underlying studies should be interpreted in light of their specific context or might be insufficiently powered, as well as noting a risk of bias in meta-analyses[10].

Against this background, the discussion centred on how AI could be used to facilitate feedback practice, raising questions about the relationship between method and technology referring to the Clark-Kozma debate concerning whether media or technology exerts effects on learning in its own right or only through the instructional methods with which they are used[23-26]. Participants emphasized that the instructional method should remain the driver of student competence development, with AI serving as a facilitator. Critical questions were: if a method is effective without AI, what additional benefits does AI provide? And how can we use AI to enhance feedback in ways that would not be possible without AI?

Addressing these questions, participants identified several ways AI could strengthen feedback practices. One of the most significant promises discussed was AI's potential to provide complex feedback more rapidly, more frequently, and on a larger scale[27]. Thereby, an advantage of AI-generated feedback would be efficiency. AI could reduce the cost and effort of generating and improving content for intelligent tutoring systems[28-30], including feedback[6,31] and free up educators from routine grading tasks[32-34]. Participants discussed how this saved time could be reallocated, ideally toward supporting students in engaging more meaningfully with the feedback. At the same time, human oversight was considered indispensable. Participants highlighted the importance of educators being involved in the feedback process in meaningful ways, enhancing and not just replacing human capability. Teachers should not only review AI feedback but also actively participate in its creation, for example, through innovative tools designed to support human-AI co-design [35-37].

A primary technological affordance of AI might be its capacity to integrate and unify previously siloed feedback functions into a single, powerful system. In the past, effective strategies like the immediate, step-by-step guidance of Intelligent Tutoring Systems, the application of psychological models of motivation, or the real-time analysis of student engagement were often confined to separate research domains or specialized tools. The new affordance of generative AI is its potential to synthesize these elements. This promises the delivery of dynamic, in-process feedback that can be continuously personalized through sophisticated learner models tracking a student's progress, cognitive state, and engagement with a specific task. Participants described this promise as AI acting like a "Thermomix" for feedback: just as the kitchen device can cook, bake, and stir, AI might potentially offer a universal architecture capable of making the best feedback strategies available to every learner, warranting careful evaluation.

Alongside these potential benefits, participants cautioned against possible detrimental effects[38]. While AI-generated scaffolding could support cognitive and metacognitive processes, overreliance on such scaffolds risks offloading these processes to the AI, which might undermine students' self-regulated learning skills and long-term learning gains [26, 38-40]. The lack of interpretability of some types of AI-generated feedback, for example, explaining why a particular piece of feedback is given, was also highlighted as a concern, as it may shape how learners and teachers respond to and whether they trust the feedback.

Related to these concerns, participants stressed the need for targeted interventions that foster AI and feedback literacy among students and teachers, as the use of Continuing Professional Development (CPD) on AI Literacy for teachers can indeed lead to increased trust and adoption[41]. Learners should be equipped to critically evaluate the limitations of AI systems, e.g., the occurrence of errors due to hallucinations in AI-generated feedback[42]. Participants stressed that such competencies are likely essential to ensure that AI feedback supports, rather than detracts from, the learning process.

**(2) Feedback Engagement and Use, and Feedback Perceptions**

A recurring theme in the discussions was the importance of students' interaction with the feedback, that is, their engagement with it, their use of it as key aspects of feedback effectiveness, as well as their feedback perceptions. Feedback engagement refers to the learner's active cognitive, behavioral, and/or affective involvement with feedback[11,43]. Engagement captures what the learner does with feedback in the moment: noticing it, allocating attention, evaluating its relevance, and deciding whether and how to incorporate it. Feedback use refers to the subsequent application of feedback to regulate or improve performance[1,44] encompassing activities such as revising work, adapting strategies, or transferring insights to future tasks. Students' engagement with feedback and use of it was widely recognized as essential, but not sufficient for feedback to be effective. Participants noted that many students do not engage with the feedback they receive[45], and even among those who do, not all benefit equally from its use[46-51]. Why some students benefit, and others do not, remains a topic for future research[52], in line with the need for research about individual differences in feedback effects. Participants called for more compelling evidence on what determines effective engagement with feedback, that is, how students can be supported in engaging with feedback and benefiting from it[49]. They suggested that research should examine multi-dimensional influences, including characteristics of the feedback itself (e.g., affective, cognitive[53], the learning content and context, and the learner[54].

The role of student perceptions of feedback in predicting feedback use was discussed. On the one hand, existing literature agrees that feedback cannot be effective unless it is used and argues that use depends partly on how students perceive the feedback and their general attitudes toward it[52, 55-57]. Recent studies suggest that students' evaluations of feedback quality might be influenced by their perceptions of the feedback provider's credibility rather than the objective quality of the feedback itself[25,43,58-59], potentially impacting AI feedback use and effectiveness. On the other hand, there is a lack of evidence on the predictive value of perceptions for actual engagement, performance gains, or competence development[60], highlighting the need for future research on this topic, examining the conditions in which feedback perceptions predict feedback use and effectiveness.

The availability of AI tools could offer new opportunities to study and optimize how learners perceive feedback and how they engage with it. By systematically varying feedback features, such as tone, timing, and level of detail, AI can be used experimentally to examine how design choices shape learners' trust, acceptance, and subsequent engagement. Developing an evidence-based understanding of how to effectively use AI to design feedback that fosters appropriate trust and genuine engagement was seen as essential to realizing its potential.

Importantly, participants emphasized that AI should not be used only to deliver feedback, but also to foster students' understanding of and engagement with feedback. This seemed crucial, especially in the context of negative feedback. One of the studies presented showed how negatively valenced feedback might backfire: it reduced self-efficacy, increased feelings of threat, and was not used by students. Based on such findings, it was also discussed how critical feedback can be designed to support student motivation and engagement, with AI providing new affordances to help students process critical feedback while maintaining their engagement. Beyond written text, AI could support the delivery of multimodal feedback, such as video or audio, potentially enhancing accessibility and engagement for diverse learners. Another promising research line deals with the role that AI can play in supporting peer-feedback for receiving as well as providing feedback [61] and supporting matching. Discussions centered on the key design factors that make feedback effective in fostering and sustaining student engagement. One example was praise and positive feedback, where evidence remains mixed depending on the type of praise, feedback provider, context, and learner characteristics [20-22, 62-63]. Drawing on motivational research, praise can enhance self-concept, motivation, and goal commitment when it recognizes controllable factors such as effort or strategy, or signals progress [64-66]. However, it can also reduce engagement when it implies that a goal has been fully achieved[9,66]. Empirical studies in educational settings reflect these contrasts: although praise often accompanies critical comments in feedback practice, it has sometimes been linked to reduced effort and performance [62], leaving open key questions about when positive feedback supports or hinders learning.

Participants discussed how AI feedback could both mitigate and amplify these challenges. On the one hand, AI can generate consistent and personalized positive feedback at scale, potentially ensuring that praise targets effort, strategy, or improvement more systematically than human feedback often does. On the other hand, the meaning of praise may change when it comes from a non-human source: learners might perceive AI praise as less authentic, altering its motivational impact. This raises new questions about how to design AI feedback that conveys encouragement without losing credibility or fostering complacency, and how learners' individual differences influence their reactions to such feedback.

In the context of discussing engagement with feedback, participants also raised methodological concerns about how engagement and perceptions are typically measured. Participants encouraged more innovative approaches to measurement that draw on motivational theories [67] to explain better why students choose (not) to engage with feedback or choose to engage with certain types of feedback and how this might relate to students' individual differences. Reliance on self-report was seen as problematic, and participants highlighted the promise of learning analytics methods (e.g., process and log data analysis) to capture when and how students engage with feedback[53, 68-70]. Insights on feedback engagement based on log data (e.g., time-on-task; usage patterns) could also be communicated back to teachers, for instance, through teacher-oriented dashboards, to support timely intervention[71-72]. Such innovative methods would allow observing student behavior in response to feedback without interfering in the learning process, and at the same time, provide interventions that could potentially support students in engaging with and learning from feedback.

### (3) Delivering Feedback at Scale

A central theme of the discussion was how to bring effective feedback into educational practice at scale. Intelligent tutoring systems were noted as established examples of feedback being delivered effectively at a large scale[73-77]. Participants noted that while a solid body of evidence is already available, mainly from U.S. secondary and higher education contexts[75] and increasingly also from European schools[78-80], further scaled trials and validation in authentic classroom settings are needed to strengthen the empirical foundation for broader implementation. Participants agreed that any scaled-up implementation should be pursued in close collaboration with practitioners and possibly commercial enterprises to increase the likelihood of successful uptake. They cautioned, however, that interventions often produce different effects in practice than in laboratory studies[79-80]. Scaling up should therefore only be done together with rigorous evaluation[81]. Complementing this, participants stressed the importance of supporting teachers with the practical aspects of implementation and the pedagogical integration of educational technologies[82-86].

Successful implementation requires not only curricular alignment but also the active involvement of teachers, not only in software design but also in classroom use and evaluation[87-89]. Such involvement helps ensure that technology complements, rather than replaces, pedagogy and remains adaptable to diverse instructional contexts. Participants discussed whether research should primarily respond to teachers' immediate needs or aim to advance beyond existing classroom practices. While the inclusion of practitioners in research was widely valued, participants noted that learners, and sometimes teachers, may not always frame their preferences in ways that align with learning sciences evidence or established pedagogical principles[90]. This highlights the need for design processes that meaningfully integrate practitioner and researcher expertise[91], with subject-specific education researchers potentially playing a mediating role. At the same time, participants emphasized that teachers cannot, and should not, be expected to master domains such as cognitive or educational psychology, prompting reflection on how co-creation with practitioners is most beneficial for both research and practice.

Related to scaling and implementation of effective feedback in authentic contexts, many discussions revolved around the contextual and individual factors of feedback, for example, the role of factors such as gender, age, (individual differences in) developmental competencies in different areas and prior (domain) knowledge of the learner population, motivational and self-regulatory requirements of the learning context, and domain-specificity, that has not yet been considered adequately in the theoretical and empirical feedback literature [92-93]. There were calls for examining feedback systematically in context (e.g., course conditions, content, feedback frequency) to move forward with understanding the determinants of feedback effectiveness, and considering different learning outcomes simultaneously (e.g., performance, motivation, emotion, meta-cognition) to understand better holistic feedback effects[42, 94-96]. To address these issues, there were calls for running an international, high-powered, preregistered, multi-site, multi-arm randomized controlled trial to provide more compelling empirical evidence to identify contextual influences and distinguish differential effects of feedback across educational contexts to understand its dependencies and boundary conditions[97-98].

Participants highlighted that AI-based systems could be key in achieving these aims by enabling large-scale, adaptive, and data-driven feedback delivery across diverse contexts. Through continuous data collection and testing, AI tools could support more systematic investigations of contextual and individual factors influencing feedback effectiveness and tailor feedback to learners' prior knowledge, motivational profiles, or developmental stages. At the same time, participants cautioned that AI may also introduce new contextual dependencies, for example, through biases in training data, limitations in domain adaptation, or mismatches between algorithmic feedback and local classroom practices. Understanding how AI feedback interacts with these contextual and individual variables, therefore, becomes central to advancing both theory and implementation of effective feedback at scale.

The potential for collaboration between research and commercial educational technology providers was discussed as a means to bring effective feedback to scale in educational practice[99-101]. However, critical issues were also raised, with key questions including: Who owns and has access to learner data? Can researchers be guaranteed continued access to evaluate effectiveness over time if companies provide the technology? Possible solutions that were discussed were the design and systematic deployment of free[35] and open-source platforms[102].

**Conclusion**

Addressing the challenges and opportunities of AI feedback will require sustained collaboration across research disciplines. Participants emphasized that the scope and complexity of these issues exceed the capacity of individual researchers or groups, calling for more systematic interdisciplinary collaboration and focused meetings to build an effective network. They highlighted the importance of establishing mechanisms that allow for more rapid, large-scale, and systematic testing of feedback effectiveness. One promising avenue discussed was coordinated large-scale experimentation, catalyzed by hypotheses submitted from the field and deployed through open-source platforms[102].

Utilizing such platforms would lower resource demands, accelerate empirical testing, and enable the synthesis of best practices. Joint efforts should prioritize longitudinal and context-specific studies, which are critical to identifying effective designs and advancing theory on the boundary conditions of feedback in AI-enhanced learning. Ensuring that AI amplifies, rather than dilutes, instructional feedback will require intentional design, robust theoretical grounding, and cumulative evidence across diverse educational contexts.

**Acknowledgments**
The symposium was supported by the Jacobs Foundation, the Leibniz Institute for Science and Mathematics Education in Kiel, Germany, and by the German Research Foundation. Jennifer Meyer is supported by a Jacobs Foundation Research Fellowship (2024-2026).

**Author contributions**
J.M., O.K., T.J., and J.F. organized the meeting. J.M. wrote the first version of the report. All authors contributed to the revision of the manuscript and read and approved the final version.

**Competing Interests**
We declare no competing interests

Figure 1.

*Overview of the Main Discussion Points*

|  | **1 Determinants of Feedback Effectiveness** | **2 Feedback Engagement and Use, and Feedback Perceptions** | **3 Delivering Feedback at Scale** |
|---|---|---|---|
| General points | • The need for theoretical integration vs. the value of specific, testable frameworks<br>• Relying on meta-analytic results vs. the need for interpreting findings in context | • The importance of student engagement: what determines effective engagement and how can we foster it?<br>• What role do student perceptions play in feedback effectiveness? | • More validation of feedback in authentic contexts is needed with focus on pedagogical integration<br>• Integration of research and practice – and the role of subject didactics as a bridge<br>• The complexities of context and learner individual differences need to be investigated to understand the boundary conditions of feedback effectiveness |
| The role of AI | • AI as facilitator of the instructional method: what added value can AI provide?<br>• The promises: Facilitation of scalable, cost-efficient, and enhanced feedback<br>• But there are risks of cognitive offloading & overreliance<br>• Mitigating the risks, there is a need for AI literacy interventions | • AI can support innovative log-based measurement of student engagement<br>• AI can support engagement feedback and not just provide it<br>• But AI might reduce feedback engagement if perceived as less credible | • AI can facilitate answering these questions, enabling large-scale investigations in authentic settings<br>• Collaboration between commercial EdTech and research, and the potential of open-source platforms with feedback facilitated by AI technology<br>• But AI feedback provides its own context and specifics (e.g., bias, trust) |